\documentclass[twocolumn,amsmath,amssymb]{revtex4}
\usepackage{bm}% bold math
\usepackage{hhline}
\usepackage{amssymb}
\usepackage{graphicx}

\begin{document}

%\preprint{APS/123-QED}

\title{The origin of the red luminescence in Mg-doped GaN}
\author{S.~Zeng}
\author{G.~N.~Aliev}
\author{D. Wolverson}
\email{d.wolverson@bath.ac.uk}
\author{J.~J.~Davies}
\author{S.~J.~Bingham}
\author{D. A.~Abdulmalik}
\author{P. G.~Coleman}
\affiliation{Department
of Physics, University of Bath, Bath, BA2 7AY, UK}
\author{T.~Wang}
\author{P.~J.~Parbrook}
\affiliation{National Centre for III-V Technologies, Department of
Electronic and Electrical Engineering, University of Sheffield,
Mappin Street, Sheffield S1 3JD, UK}

\date{\today}

\begin{abstract}
Optically-detected magnetic resonance (ODMR) and positron
annihilation spectroscopy (PAS) experiments have been employed to
study magnesium-doped GaN layers grown by metal-organic vapor phase
epitaxy. As the Mg doping level is changed, the combined experiments
reveal a strong correlation between the vacancy concentrations and
the intensity of the red photoluminescence band at 1.8 eV. The
analysis provides strong evidence that the emission is due to
recombination in which electrons both from  effective mass donors
and from deeper donors recombine with deep centers, the deep centers
being vacancy-related defects.
\end{abstract}


\maketitle

Deep defects play a key role in the performance limits and aging
effects of GaN-based light-emitting devices. They also lead to
photoluminescence (PL) at energies well below the band-gap. For
example, PL and optically-detected magnetic resonance (ODMR) studies
\cite{Kaufmann99, Hofmann00, Bayerl01, Bayerl99pss} have suggested
that deep defects are responsible for the red (1.8 eV) luminescence
band which is often observed in Mg-doped GaN and that the band is
due to recombination emission in which vacancy-dopant complexes are
involved \cite{Kaufmann99,Hofmann00}. However, this proposal was
mainly based on indirect evidence and on previous experience of II -
VI compounds, and further experimental confirmation is therefore
needed. The present study involved the use of both ODMR and positron
annihilation spectroscopy (PAS) on the same set of samples covering
a range of Mg doping levels and we have established a correlation
between the ODMR spectra (obtained by monitoring the red PL) and the
PAS results.

ODMR is well established as a means of investigating centers
involved in recombination processes in semiconductors~\cite{
Kennedy98, Meyer}. For a detailed description of the technique and
our ODMR system, see Ref. \cite{Aliev05}. The ODMR was carried out
at 14 GHz with the specimen at 2K. The PL was excited with a UV
argon-ion laser (363.8/351.1 nm). The microwaves were chopped at 605
Hz and changes in the PL intensity caused by magnetic resonance were
monitored at this frequency as the magnetic field was slowly swept.
PAS with a slow positron beam is an effective tool for the
investigation of open volume defects such as neutral or negatively
charged vacancies in semiconductor films. When positrons annihilate
electrons in semiconductors the resulting gamma ray energy spectrum,
peaked at 511 keV, is Doppler-broadened (since the electrons have a
range of momenta). The annihilation linewidth is characterized by
quantities $S$ ($W$), defined as the central (wing) fraction of the
line. The value of $S$ ($W$) is characteristic of the material under
study, but is generally higher (lower) when vacancies are present
\cite{R. Krause}. Measurements of $S$ ($W$) can thus be used to
monitor vacancy concentrations. In the present work, single-detector
Doppler-broadening PAS was performed using a magnetic transport
positron beam system \cite{Coleman95}. Positrons were implanted into
the layers at energies in the range 0.1 - 30 keV, corresponding to
mean depths up to 1.5 nm.

Details of the growth of the  GaN:Mg samples by metallo-organic
vapor phase epitaxy (MOVPE) were given earlier \cite{Aliev05}. The
Mg concentrations in the layers were not determined directly but are
expected to increase monotonically with the precursor flow rate,
which was 75, 100, 200 and 300 sccm for samples \#626, \#625, \#624
and \#623 respectively. A nominally undoped highly resistive sample
(\#621) was grown under the same conditions. A piece of the most
lightly doped specimen (\#626) was annealed at 850$^\circ$C in a
$\rm{N_2}$ atmosphere in which the $\rm{O_2}$ concentration was
0.5\%.  The majority of experiments reported below were on the {\it
as-grown} material.

\begin{figure}[t]
\centerline{\includegraphics[keepaspectratio=true,scale=0.3]{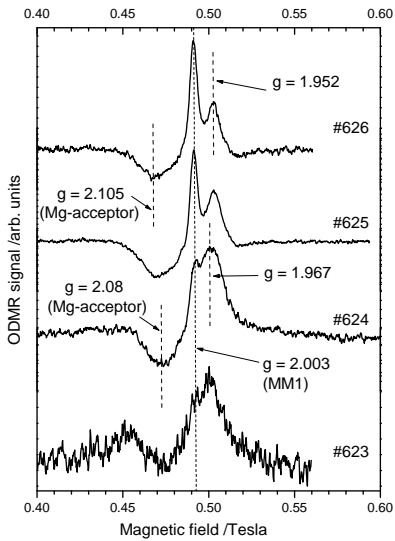}}\caption{\label{figure1}
ODMR spectra detected from the 1.77 - 1.91 eV spectral region for
samples \#626, \#625, \#624, and \#623 (in order of increasing Mg
concentration). The arrows indicate the \emph{g}-values at ${\bf
B}\|c$ obtained using a Lorentzian multipeak fitting procedure and
correspond to the resonances indicated by vertical lines.}
\end{figure}

For the Mg-doped samples, we observed a broad red luminescence band
peaked at $\sim 1.8$ eV. It is strongest in the lightly doped
material, becomes weaker as the Mg concentration is increased and
vanishes after annealing. The ODMR signals detected from the 1.77 -
1.91 eV spectral region of the four as-grown Mg-doped samples are
shown in Fig.~\ref{figure1}. Both PL-enhancing and PL-quenching
signals are observed. The sharp enhancing signal in the middle of
the spectra is isotropic with $g_\|, g_\bot=2.003\pm0.003$ and FWHM
= 5 mT. The \emph{g}-value and width correspond closely to those of
the so-called MM1 center \cite{Bayerl01, Bayerl99pss}: this signal
is PL-enhancing only when detected via the red band and has been
attributed \cite{Bayerl01, Bayerl99pss} to deep defects in the lower
midgap region. At the high field side of the sharp signal two other,
overlapping, luminescence-enhancing resonances are observed, which
are better resolved in measurements performed at 34 GHz
\cite{Gazi04}. From comparison with previous magnetic resonance work
\cite{Bayerl01,Glaser02, Carlos93, Glaser99}, the narrower
resonance, with $g_\|=1.952\pm0.003$ and FWHM = 8 - 10 mT, is
attributed to effective-mass (EM) donors. The \emph{g}-value of the
broader signal ($g_\|=1.967\pm0.005$) is close to that of a deeper
donor signal ($g_\|=1.962$) detected from the 2.75 - 3.1 eV spectral
region of both the as-grown and annealed samples that we have
studied here \cite{Gazi04}, suggesting that the same deeper donor is
associated with both the 2.8 eV ``blue'' band and the 1.8 eV ``red''
band. In addition to the PL-enhancing lines, an asymmetric quenching
signal with a \emph{g}-value of about 2.105 is detected, and is
attributed to shallow Mg acceptors \cite{Glaser02}: its PL-quenching
character indicates that it is a center involved in a recombination
process competing with the red emission. The ODMR results presented
above suggest that donor (EM and/or deeper)-to-deep-center
recombination is responsible for the red emission. As in the PL
spectra, the ODMR signals in the red become weaker as the Mg
concentration is increased. The signals are absent from undoped
samples and, in doped samples, are suppressed after annealing.

\begin{figure}[t]

\centerline{\includegraphics[keepaspectratio=true,scale=0.36]{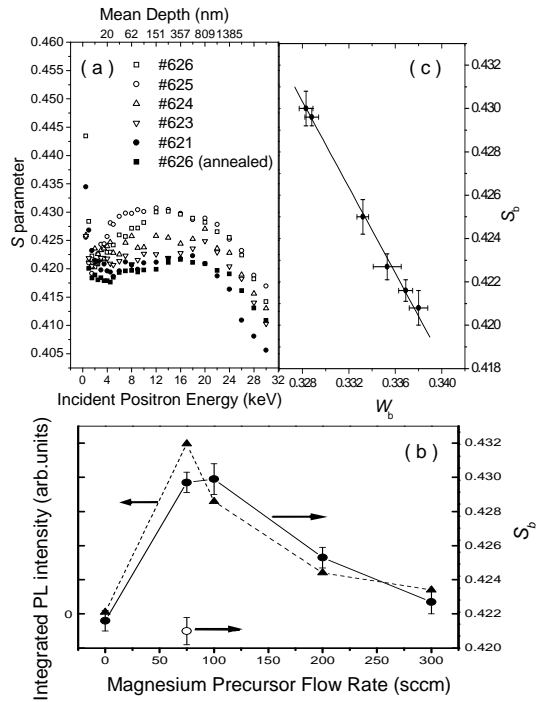}}\caption{\label{figure2}
(a) The low momentum $S$ parameter versus incident positron energy
for the range of as-grown and annealed GaN samples; (b) The change
of the mid-range $S_b$ parameter with Mg concentration (filled
circles) and the behavior of the integrated intensity of the red
emission (filled triangles). The unfilled circle shows the
mid-range $S_b$ parameter for the most lightly doped specimen
after annealing. The lines are guides to the eye; (c) $S_b$ versus
$W_b$ for the samples studied in the present work.}
\end{figure}

The PAS results are presented in Fig.~\ref{figure2}(a). The $S$
parameters of the Mg-doped as-grown samples, along with those of the
annealed sample and of the undoped reference sample, are shown as a
function of positron implantation energy. The corresponding mean
depths probed are also shown. For the undoped and the annealed
samples, as the positron implantation energy increased, the $S$
parameters firstly decrease from the surface specific value (i.e.,
at low energy) and become approximately constant at around 0.420
above 2.5 keV, when almost all positrons are annihilated in the GaN
film. The slight rise in $S$ parameters between 2.5 keV to 18 keV is
consistent with a previous report \cite{Xiaodong02} which suggested
that vacancies prefer to reside alongside dislocations, whose
density increases towards the interface of the film and the
nucleation layer. Beyond 18 keV, $S$ falls towards the value
characteristic of the sapphire substrate (measured separately to be
0.392). Fig.~\ref{figure2}(b) shows $S_b$ as a function of Mg
concentration, where $S_b$ ($W_b$) is taken as the average of the
$S$ ($W$) parameter measured in the energy range from 12 keV to 18
keV and represents the bulk film value. The highest $S_b$ values are
observed for the most lightly-doped as-grown samples. With
increasing Mg concentration, the $S_b$ value decreases and
approaches those of the undoped and of the annealed samples. The
dependence of $S_b$ on doping and annealing is very similar to that
of the red luminescence (shown also in Fig.~\ref{figure2}(b)) and of
the ODMR signals detected in the red spectral region. This
correlation thus provides strong support to the assertion that the
red emission is associated with vacancy-type defects. Furthermore,
the linear relationship between $S_b$ and $W_b$ shown in
Fig.~\ref{figure2}(c) is consistent with the existence of only one
type of vacancy-related defect in the films \cite{R. Krause}. The
evidence points to the defects being related to the deep center
signal in the ODMR spectra. In a separate ODMR experiment on the 2.8
eV ``blue" band, the deeper donor signal ($g_\|=1.962$) was obtained
in both as-grown and annealed samples; therefore, it is unlikely
that the donors account for the changes of the spectra.

We turn now to the identity of the deep vacancy-related center. Both
$\rm{V_{Ga}}$ \cite{Saarinen97} and $\rm{V_N-Mg_{Ga}}$
\cite{Hautakangas03} have been observed in GaN by PAS. $\rm{V_{Ga}}$
would be expected to be negatively charged and thus to act as a
positron trap. Calculations\cite{Mattila97, Van de Walle94} have
shown that the formation energy of $\rm{V_{Ga}}$ is increased as the
Fermi level moves towards the valence band. This result can be used
to explain the decrease of strength of the red emission with Mg
doping and annealing, for the reason that the concentration of
$\rm{V_{Ga}}$ decreases. To be observed in the ODMR spectrum, the
vacancy would be expected to contain an odd number of holes in the
optically-excited magnetic state and thus to be $\rm{V_{Ga}^{0}}$ or
$\rm{V_{Ga}^{2-}}$ (in the Sonder-Sibley\cite{Sonder} notation, we
write these as $[3{\rm h}^{+}]_{\rm Ga}$ and $[{\rm h}^{+}]_{\rm
Ga}^{\rm {2-}}$ respectively). The $g$-value would be expected to be
close to 2.00. It is also possible that the deep center is formed
from an association between a Ga vacancy and an oxygen ion \cite
{Reshchikov02}.

However, $\rm{V_N-Mg_{Ga}}$ dominates over $\rm{V_{Ga}}$ in Mg-doped
GaN grown via MOVPE \cite{Hautakangas03}. If this complex forms the
present deep center, the PAS results require it to be neutral or
negatively charged (in order to trap the positrons). The neutral
state, in the Sonder-Sibley notation, can be written as
$([2e^{-}]_{N}^{3+}+ [\rm {Mg}^{2+}]_{Ga}^{-})^{0}$, where we have
assumed the electron to be trapped in the vacancy. Under optical
excitation the complex may lose an electron and become paramagnetic,
leading again to an ODMR signal close to 2.00. It has been reported
that the complex dissociates under $p$-type conditions
\cite{Hautakangas03} and therefore, with Mg doping and annealing,
its concentration would be expected to decrease, leading to the
decrease of the $S$ parameter and the red emission.

\begin{figure}[t]
\centerline{\includegraphics[keepaspectratio=true, clip=true,
viewport=135 465 416 698,
height=2in]{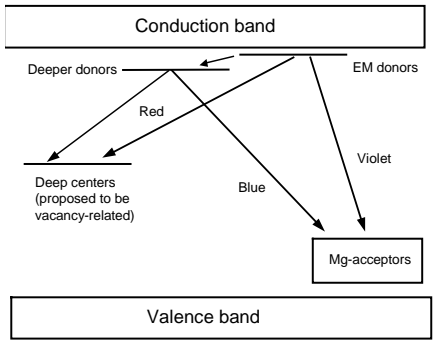}}\caption{\label{figure3} Schematic diagram
for the recombination process giving rise to the red, blue and
violet bands of GaN.}
\end{figure}

There are thus two possible models for the deep center, one
involving gallium vacancies and the other nitrogen vacancy
complexes. A one-electron energy level diagram which summarizes the
conclusions is shown in Fig.~\ref{figure3}. The diagram is similar
to those in Refs. \cite{Kaufmann99} and \cite{Bayerl01}, except that
the difference in the energies of the EM and deeper donor states in
our case needs to be much smaller than the 250 meV proposed in the
former reference (this is since, in our case, the emission bands
involving either the EM and or the deeper donors both lead to red
emission). If we take the MM1 center to be Ga vacancy-related
center, the scheme of Fig.~\ref{figure3} is in agreement with that
shown in Fig. 5 of Ref. \cite{Bayerl01} (but in contrast to the
assignment in Ref. \cite {Hofmann00}). Fig.~\ref{figure3} is almost
certainly oversimplified and should be viewed with caution, since
energy levels other than those shown are likely to be present.
Further, there is strong evidence that the depth of the {\it
shallow} acceptors (presumed to be formed by substitutional
magnesium at gallium sites) is affected by Jahn-Teller effects and
by the influence of strain and other nearby defects \cite{Aliev05},
so that it is represented in the diagram by a range of energy
levels.

In summary, epitaxial GaN layers have been studied using ODMR and
PAS. We find that the change of the PAS parameter $S$ with Mg
concentration and annealing is correlated with the behavior of the
PL and ODMR signals in the red region, indicating that the red
luminescence is vacancy-related. The experiments lead to the
conclusion that the red emission is due to recombination between
electrons both from  EM and from deeper donors with deep centers and
point to the deep centers being vacancy-related defects.

We are grateful for support from the EPSRC (project GR/R34066) and
for scholarships from Universities UK (SZ) and the State of Qatar
(DAA). We thank R.J. Lynch for temperature treatment of the samples,
S. Stepanov for Hall-effect measurements and X. D. Pi for valuable
discussions.

\end{document}